# Magnetic and magnetoelectric studies in pure and cation doped BiFeO$_3$

V. B. Naik and R. Mahendiran

Department of Physics and NUS Nanoscience and Nanotechnology Initiative (NUSNNI),

Faculty of Science, 2 Science Drive 3,

National University of Singapore - 117 542, Singapore

Abstract

We report the effect of divalent cation (A) substitution on magnetic and magnetoelectric properties in Bi$_{1-x}$A$_x$FeO$_3$ (A= Sr, Ba and Sr$_{0.5}$Ba$_{0.5}$; x = 0 and 0.3). The rapid increase of magnetization below 100 K and a peak at the Neel temperature $T_N$ = 642±2 K found in BiFeO$_3$ is suppressed in the co-doped sample (A = Sr$_{0.15}$Ba$_{0.15}$). All the divalent cation doped samples show enhanced magnetization with a well defined hysteresis loop compared to the parent compound. Both longitudinal (L-$\alpha_{ME}$) and transverse (T-$\alpha_{ME}$) magnetoelectric coefficients with *dc* magnetic field parallel and perpendicular to the direction of induced voltage, respectively, were measured using dynamic lock-in technique. It is found that the T-$\alpha_{ME}$ increases in magnitude and exceeds the L-$\alpha_{ME}$ with increasing size of the A cation. The maximum T-$\alpha_{ME}$ = 2.1 mV/cmOe in the series is found for A = Sr$_{0.15}$Ba$_{0.15}$, though it is not the compound with the highest saturation magnetization. The observed changes in the magnetoelectric coefficient are suggested to possible modification in the domain structure and magnetoelectric coupling in these compounds.

## I. INTRODUCTION

Multiferroic materials that are simultaneously ferromagnetic and ferroelectric are currently attracting great attention because the possibility of modulating electrical polarization with a magnetic field and magnetization with an electric field in these materials can pay away new technologies which will exploit both electrical and magnetic polarizations to store and manipulate information[1,2,3,4]. In addition, magnetic control of dielectric permittivity in multiferroic oxides can be used to create new kind of high frequency filters and wireless sensors.[5,6] However, single phase multiferroic materials are rare since ferroelectricity in $ABO_3$ type perovskite oxide requires transition metal ions (B) with $d^0$ electronic configuration, whereas ferromagnetism requires transition metal ions with odd number of $d$ electrons. This apparent incompatibility can be overcome in materials such as $BiFeO_3$, $PbFe_{0.5}Nb_{0.5}O_3$ and $ABO_3$ (A = Y, Tb, Gd, Ho; B=Mn)[7], where A and B site cations are sources of ferroelectricity and magnetism, respectively. Though the physics of ferroelectricity and magnetoelectric (*ME*) coupling in $AMnO_3$ is very exciting, they are not useful for practical applications at room temperature since the magnetic field-induced ferroelectric polarization is rather small and the magnetic transition occurs mostly below the liquid nitrogen temperature. The ferrimagnetic magnetic transition of $PbFe_{0.5}Nb_{0.5}O_3$ is also below 300 K.[8]

The high values of ferroelectric Curie temperature ($T_{C(FE)}$ ~ 836 °C)[9,10] and antiferromagnetic transition temperature ($T_N$ ~ 370 °C)[11] found in $BiFeO_3$ make this compound more attractive. Ferroelectricity in $BiFeO_3$ is driven by the sterochemical activity of $6s^2$ lone pair of $Bi^{3+}$ ions, whereas the Fe ions order antiferromagnetically. It is a G-type antiferromagnet with a long-range cycloidal spin arrangement of wavelength 62 nm incommensurate with the lattice.[12] The saturation polarization in this material can reach as much as 60 $\mu C/cm^2$ in suitably prepared thin films,[13] though it is generally small in bulk sample.[9,14] However, ferro or ferrimagnetism instead of antiferromagnetism along with enhanced polarization is highly preferable for low-field practical applications. Enhancements in magnetization and ferroelectric polarization were reported in the A-site doped system of $Bi_{0.9-x}Tb_xLa_{0.1}FeO_3$ where Tb and La are isovalent to $Bi^{3+}$ ions[15], and also in $Bi_{1-x}Nd_xFeO_3$.[16] Interestingly, divalent cation (A) substituted $Bi_{0.7}A_{0.3}FeO_3$ (A = Ca, Sr, Pb, and Ba) also exhibits enhanced magnetization[17] and the recent observation of magnetic-field induced ferroelectric hysteresis loop in $Bi_{0.75}Sr_{0.25}FeO_{3-\delta}$ makes it more attractive for practical applications.[18] Although the temperature dependence of structural and magnetic order parameters of $BiFeO_3$ has been studied by Fischer *et al* long ago[19], the magnetic measurements over a wide range of temperature (10 K – 700 K) as well as magnetoelectric coupling in $Bi_{0.7}A_{0.3}FeO_3$ have not been reported and these are the driving factors for the work reported here.

## II. EXPERIMENT

Polycrystalline samples of $BiFeO_3$ and $Bi_{0.7}A_{0.3}FeO_3$ (A = Sr, Ba and $Sr_{0.5}Ba_{0.5}$) were synthesized by the conventional solid state reaction. The solid solutions were prepared by mixing and grinding stoichiometric mixtures of $Bi_2O_3$, $SrCO_3$, $BaCO_3$ and $Fe_2O_3$ in agate mortar and preheating the powder in the temperature range from 800 °C (x = 0 )- 850 °C (x=0.3) for 5-6 hours. Structural and phase characterizations were done with X-ray diffraction (XRD) experiment (Philips X' pert Pro) using Cu Kα radiation. All the compounds were pressed into pellets and final heat treatment was done in the temperature range from 850 °C (pure BiFeO$_3$) to 925 °C (Sr) for 5 hours to get relatively dense pellets. For magnetoelectric measurements, all the samples were cut into a rectangular shape of similar dimensions, polished and made into parallel plate capacitor structure by applying silver paint on opposite faces of the sample and heat treated. The ferroelectric properties (*P-E* loop) were measured using a Precision LC ferroelectric tester (Radiant Technologies). The temperature and field dependence of magnetization were measured using a vibrating sample magnetometer (VSM) with a superconducting magnet (Quantum Design, USA) in the temperature range $T$ = 10 - 400 K, and the high temperature ($T$ = 300 K – 700 K) dependence of magnetization under a fixed magnetic field ($H_{dc}$ = 0.5 T) was done using a VSM with an electromagnet (Lakeshore – model: 74034).

Figure 1 shows the schematic diagram of the dynamic lock-in technique used to measure the magnetoelectric effect which is similar to the experimental set up used by Duong *et al.*[20] The sine output of the internal function generator of a lock-in amplifier (Stanford Research, model SR850) was given to a power amplifier (PA) and the amplified current fed the Helmholtz coils (HC – 130 turns of AWG 26 copper wire with a diameter of 40 mm) which provided *ac* magnetic fields up to $h_0$ = 10 Oe at a frequency of 7 kHz. We restricted our measurements to the optimal frequency, *f* = 7 kHz due to the limitation of the power amplifier (the gain of the amplifier decreases for frequency above 10 kHz, and hence the magnitude of the *ac* magnetic field decreases). The *ac* magnetic field at the centre of the Helmholtz coils was monitored using Hall probe (HP) with Gauss meter (GM). In addition, the strength of the magnetic field was estimated by measuring the *ac* voltage developed across a standard resistor connected in series with the Helmholtz coils using a Keithley 2700 digital multimeter. A Yokogawa GS610 – source measure unit (SMU) was used to ramp *dc* current through the electromagnet (EM) up to ±3 A to provide a maximum field of ± 3 kOe with the pole spacing of 20 mm.

The reorientation of the electrical dipoles in the sample by *ac* magnetic field induces *ac* voltage on the top and bottom surfaces of the sample through magnetoelectric coupling, which was measured using the lock-in amplifier in the differential mode to eliminate the errors due to induction by Faraday effect. The *ME* coefficient, $\alpha_{ME}$ can be calculated using relation $\alpha_{ME} = \frac{dE}{dH} = \frac{1}{t}\frac{dV}{dH} = \frac{V_{out}}{h_0 t}$ where, $V_{out}$ is the *ac* magnetoelectric voltage appeared across the sample surface (measured by the lock-in), $h_0$ is the amplitude of the *ac* magnetic field, and *t* is the thickness of the sample. Both longitudinal and transverse magnetoelectric coefficients were measured. The longitudinal magnetoelectric coefficient refers to the measurement of magnetoelectric voltage parallel to the direction of the applied magnetic fields (*ac* and *dc*) i.e. the silver paste coated surfaces of the sample is perpendicular to the magnetic fields. In the transverse case, silver paste coated surfaces of the sample is parallel to the magnetic fields. All the instruments were controlled by the LabVIEW 8.2 software programme.

### III. RESULTS AND DISCUSSION

Figure 2 [fig. 2(a) to 2(d)] shows the X-ray diffraction pattern of all the samples at room temperature. The BiFeO$_3$ shows a small fraction of impurity phases such as Bi$_2$Fe$_4$O$_9$, Bi$_{25}$FeO$_{40}$ [marked by * in fig. 2(a)] in addition to the main rhombohedral

phase as reported earlier.[21] It has been known that single phase BiFeO$_3$ is very difficult to prepare by the standard solid state chemistry route. However, impurity phases almost disappear in the Sr and Ba doped BiFeO$_3$ compounds as it can be seen in fig. 2(b) to fig. 2(d). The observed XRD pattern of one of the doped compounds, Sr and Ba co-doped, was indexed to rhombohedral structure of BiFeO$_3$ with space group R3c using TOPAS software version 2.1 [fig. 2(e)]. The lattice parameters of the co-doped compound are a = 5.592 Å and c = 13.809 Å. The detail structural studies were beyond the scope of this letter and will be published elsewhere.

The main panel of figure 3 shows *M(T)* measured at *H* = 0.5 T for the two selected samples, BiFeO$_3$ and Bi$_{0.7}$Sr$_{0.15}$Ba$_{0.15}$FeO$_3$ over a wide temperature range from *T* = 700 K to 10 K. The rapid increase of magnetization at $T_N$ = 644±2 K signals the onset of antiferromagnetic transition in BiFeO$_3$.[22] The $T_N$ is determined from the inflection point of the *M(T)* curve around the transition. Furthermore, *M(T)* decreases monotonically below $T_N$ and then raises rapidly below 100 K. This complex magnetization behavior is possibly due to the development of incommensurate sinusoidal spin structure in BiFeO$_3$. On the other hand, *M(T)* for Bi$_{0.7}$Sr$_{0.15}$Ba$_{0.15}$FeO$_3$ shows only a weak anomaly around $T_N$ = 642±2 K and increases gradually with lowering in temperature. The magnitude of *M* at 10 K of the latter compound is an order of magnitude higher than the BiFeO$_3$. The inset compares the *M* versus *T* at *H* = 0.5 T for all four compounds below 400 K. We see that the magnetization increases with lowering temperature for all the doped samples and the Ba (30%) doped compound showed the highest value.

The main panel of figure 4 shows the field dependence of magnetization for BiFeO$_3$ and Bi$_{0.7}$(Sr, Ba)$_{0.3}$FeO$_3$ at room temperature. The magnetization of pure BiFeO$_3$ increases linearly without saturation up to the maximum field ($H$ = 5 T) which confirms that the sample is in the antiferromagnetic state even at the highest field. However, the divalent doped compounds show a weak ferromagnetic like behavior with well developed hysteresis loop and enhanced magnetization. The spontaneous magnetization ($M_s$) obtained from extrapolation of the high field $M$ to $H$ = 0 increases with increasing size of the divalent cation ($M_s$ = 0.098 emu/g for A = Sr, and 0.75 emu/g for A = Ba). Though coercive field initially increases from $H_c$ = 0.27 T for A = Sr to 0.46 T for A= Sr$_{0.15}$Ba$_{0.15}$, it decreases to $H_c$ = 0.32 T for A = Ba. The $M$-$H$ loops at $T$ = 10 K shown on the top-right inset indicates a slight decrease in the coercive field ($H_c$ = 0.15 T for A = Sr and 0.37 T for A= Sr$_{0.5}$Ba$_{0.5}$) compare to the room temperature values. Our $M$ versus $H$ results on A = Ba, Sr compounds are in close agreement with the recent report by V. A. Khomchenko et al.[17]

The divalent substitution in Bi site is expected to convert a fraction of Fe$^{3+}$ to Fe$^{4+}$. However, Fe$^{4+}$ is difficult to form at ambient pressure and the presence of Fe$^{4+}$ ions is not supported by the Mossbauer spectroscopy.[18] Hence, the charge imbalance introduced by divalent cations has to be compensated by oxygen deficiency (O$_{3-\delta}$ instead O$_3$ with $\delta \approx 0.15$). We note that the ionic radii of Bi$^{3+}$, Sr$^{2+}$ and Ba$^{2+}$ are 1.03 Å 1.18 Å and 1.36 Å, respectively[23]. The systematic increase in the spontaneous magnetization with increasing average ionic radii of the divalent cation possibly indicates that the oxygen deficiency alone may not cause of the enhanced magnetization, but it may be related to the progressive suppression of spiral spin structure and/or increase in the canting angle of antiferromagnetically coupled layers due to tilting of $FeO_{6-\delta}$ octahedra. However, it is difficult to draw a clear conclusion without magnetic neutron diffraction or other structural studies on these divalent doped compounds. It is interesting to note that the magnetic moment obtained with divalent cation substitution is comparable to the rare earth substituted $BiFeO_3$.[24, 25]

The *P-E* loops at room temperature for $BiFeO_3$ and $Bi_{0.7}Sr_{0.15}Ba_{0.15}FeO_3$ are shown in fig. 5. The $BiFeO_3$ exhibits an unsaturated hysteresis loop which is rounded at the highest field. Though $BiFeO_3$ is ferroelectric, the observed hysteresis loop is due to a large leakage current which can overshadow the real contribution from reorientation of electrical dipoles. On the other hand, the co-doped compound exhibits a smaller hysteresis and polarization compared to the parent compound. It suggests that the leakage current or dielectric loss is negligible in doped compound compared to the parent one.

Fig. 6 shows the magnetic field dependence of the magnetoelectric coefficient ($\alpha_{ME}$) at room temperature for (a) $BiFeO_3$ ($x = 0$), (b) A = Sr, (c) A = Ba, and (d) A = $Sr_{0.5}Ba_{0.5}$. The longitudinal (*L-$\alpha_{ME}$*) and the transverse (*T-$\alpha_{ME}$*) magnetoelectric coefficients are plotted on the right and left scales, respectively in all graphs. The measurements reported here are at a fixed frequency *f* = 7 kHz of the *ac* signal. This is the optimum frequency for the maximum gain of the power amplifier we used. All the samples exhibit hysteresis over a certain field range. For $BiFeO_3$ compound [fig. 6(a)], as the field increases from zero, the *L-$\alpha_{ME}$* increases rapidly and exhibits a sharp maximum around *H* = 0.15 kOe, where *L-$\alpha_{ME}$* = 0.6 mV/cmOe. Above *H* = 0.9 kOe, *L-$\alpha_{ME}$* varies little with the magnetic field. Upon reducing the field from the maximum value, hysteresis appears in the field range *H* = +0.9 kOe to *H* = -0.9 kOe and a peak at *H* = - 0.15 kOe. The $T\text{-}\alpha_{ME}$ shows a similar behavior, but the hysteresis region is widened and the maximum $T\text{-}\alpha_{ME} = 0.28$ mV/cmOe occurs at $H = 0.19$ kOe. The maximum $L\text{-}\alpha_{ME}$ observed in our sample is an order of magnitude smaller than earlier report on the same composition by Caicedo et al.[26] This discrepancy is possibly due to the different sample preparation conditions. The $\alpha_{ME}$ of the Sr (30%) doped $BiFeO_3$ [fig. 6(b)] shows similar behavior as $BiFeO_3$. The maximum $\alpha_{ME}$ is 0.34 mV/cmOe observed in this compounds is slightly lower than that of $BiFeO_3$, and we note that $L\text{-}\alpha_{ME}$ and $T\text{-}\alpha_{ME}$ are comparable in values. However, $L\text{-}\alpha_{ME}$ of $Bi_{0.7}Ba_{0.3}FeO_3$ [fig. 6(c)] shows a completely different behavior: it shows a less pronounced peak at low fields compared to the parent compound and decreases almost linearly with the field above $H = 0.45$ T without saturation. Hysteresis occurs over a wide-field range compared to the undoped and Sr (30%) doped compounds. The field dependence of $T\text{-}\alpha_{ME}$ is similar to the parent compound but the magnitude of its peak is nearly twice that of parent compound. The $L\text{-}\alpha_{ME}$ of the co-doped compound, $Bi_{0.7}Ba_{0.15}Sr_{0.3}FeO_3$ is larger than that of parent compound. The $L\text{-}\alpha_{ME}$ of the co-doped sample exhibits a complex hysteresis loop and showed the highest value ($\approx 2.3$ mV/cmOe at $H_{dc} = 0.25$ kOe) in the series even though this is not the compound with the highest saturation magnetization.

The above data suggest that the transverse magnetoelectric coefficient is comparable or exceeds the longitudinal magnetoelectric effect in the divalent doped compounds, and the highest magnetoelectric coefficient is obtained for the co-doped compound. The peak in the magnetoelectric coefficient does not coincide with the coercive field, instead it occurs below the coercive field. Hence, the peak in $\alpha_{ME}$ is caused by a different mechanism. It is most likely related to the inflection point of the

d$\lambda$/dH curve, where $\lambda$ is the magnetostriction.[27] The observed variations in the longitudinal and transverse magnetoelectric effects with increasing size of the A cation possibly arises from the changes in magnetic domain structure and magnetostriction in these compounds. However, further investigations such as magnetostriction as a function of magnetic field is needed to elucidate the origin of the observed effect.

## IV. CONCLUSIONS

We found that the divalent cation doping in the antiferromagnetic BiFeO$_3$ enhances the magnetization with a well developed hysteresis loop due the effective suppression of spiral spin structure, and the magnitude of the spontaneous magnetization increases with the size of the dopants. It is found that the transverse $\alpha_{ME}$ increases in magnitude and exceeds the longitudinal $\alpha_{ME}$ with increasing size of the A cation. The A = Sr$_{0.15}$Ba$_{0.15}$ shows the maximum transverse ME coefficient, T-$\alpha_{ME}$ = 2.1 mV/cmOe in the series, though it is not the compound with the highest saturation magnetization. The observed changes in the transverse ME coefficient is suggested to possible modification in the domain structure and magnetoelectric coupling. The co-doped compound, A = Sr$_{0.15}$Ba$_{0.15}$ exhibits lesser leakage current than the parent compound. Further investigations, particularly magnetostriction and electrical polarization at higher electric field studies will be helpful to understand the origin of the enhanced magnetoelectric coefficient in the divalent doped BiFeO$_3$.

**ACKNOLEDGEMENTS**

R.M acknowledges the Singapore National Research Foundation for supporting this work through the grant NRF-CRP-G-2007. The authors wish to thank Dr. M. V. V. Reddy for helping out in performing Reitveld analysis.

**Figure Captions:**

**Fig. 1 Schematic diagram of the experimental set up to measure the magnetoelectric coefficient using dynamic lock-in technique: where, SMU – source measure unit, PA – power amplifier, HC – Helmholtz coil, EM – electromagnet, $HP_{ac}$ – $HP_{dc}$, and $GM_{ac}$ – $GM_{dc}$ are the Hall probes and Gauss meters for *ac* and *dc* magnetic fields, respectively. Both Hall probes are close to each other unlike it appears in the diagram.**

**Fig. 2 XRD patterns of (a) $BiFeO_3$, (b) A = $Ba_{0.3}$ (c) A = $Sr_{0.3}$ (d) A = $Sr_{0.15}Ba_{0.15}$ compounds at room temperature (RT); (e) Reitveld refinement of the XRD pattern for the $Bi_{0.7}Sr_{0.15}Ba_{0.15}FeO_3$ compound at RT with space group R3c.**

**Fig. 3 The temperature dependences (10 K – 700 K) of magnetization at H = 5 kOe for $BiFeO_3$ and $Bi_{0.7}Sr_{0.15}Ba_{0.15}FeO_3$ compounds. The inset shows the temperature dependences of magnetization ($T$ = 10K – 400 K) at 5 kOe for all the compounds.**

**Fig. 4 Field dependences of magnetization for $Bi_{0.7}(Sr,Ba)_{0.3}FeO_3$ compounds at room temperature. The inset shows the field dependences of magnetization at 10 K.**

**Fig. 5 *P-E* loops for $BiFeO_3$ and $Bi_{0.7}Sr_{0.15}Ba_{0.15}FeO_3$ compounds at room temperature .**

**Fig. 6** Room temperature *dc* bias magnetic field dependence of longitudinal (left scale) and transverse (right scale) magnetoelectric coefficients ($\alpha_{ME}$) for BiFeO$_3$ (x = 0) and Bi$_{0.7}$(Sr,Ba)$_{0.3}$FeO$_3$ compounds in ac magnetic field of frequency $f$ = 7 kHz.

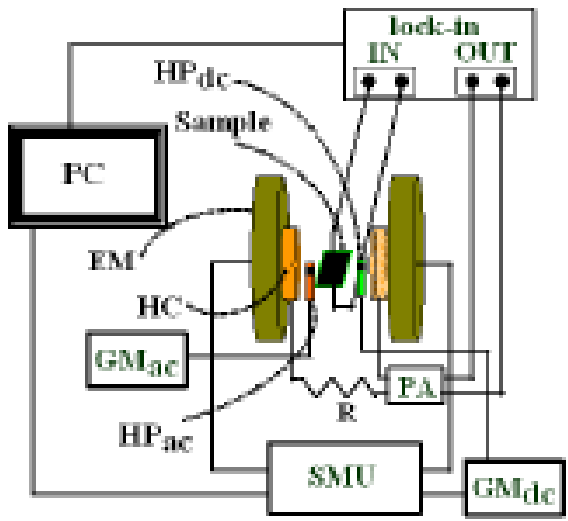

**Fig. 1 Schematic diagram of the experimental set up to measure the magnetoelectric coefficient using dynamic lock-in technique: where, SMU – source measure unit, PA – power amplifier, HC – Helmholtz coil, EM – electromagnet, $HP_{ac}$ – $HP_{dc}$, and $GM_{ac}$ – $GM_{dc}$ are the Hall probes and Gauss meters for *ac* and *dc* magnetic fields, respectively. Both Hall probes are close to each other unlike it appears in the diagram.**

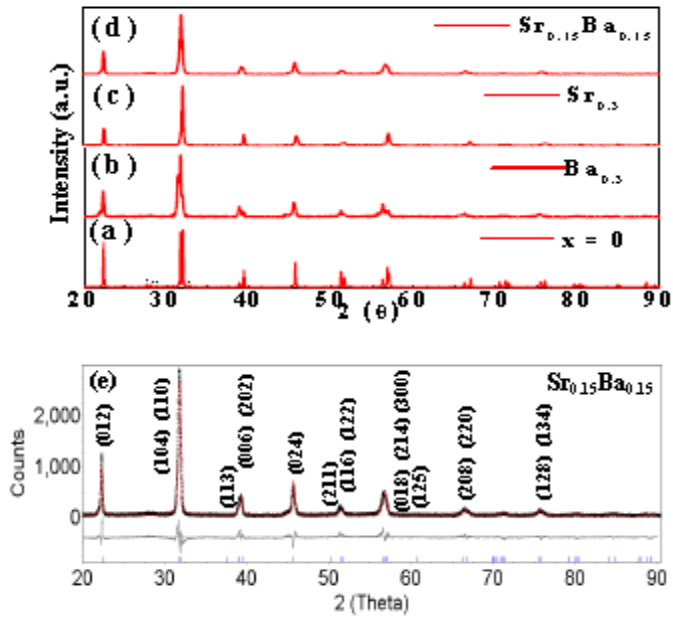

**Fig. 2** XRD patterns of (a) $BiFeO_3$, (b) A = $Ba_{0.3}$ (c) A = $Sr_{0.3}$ (d) A = $Sr_{0.15}Ba_{0.15}$ compounds at room temperature (RT); (e) Reitveld refinement of the XRD pattern for the $Bi_{0.7}Sr_{0.15}Ba_{0.15}FeO_3$ compound at RT with space group R3c.

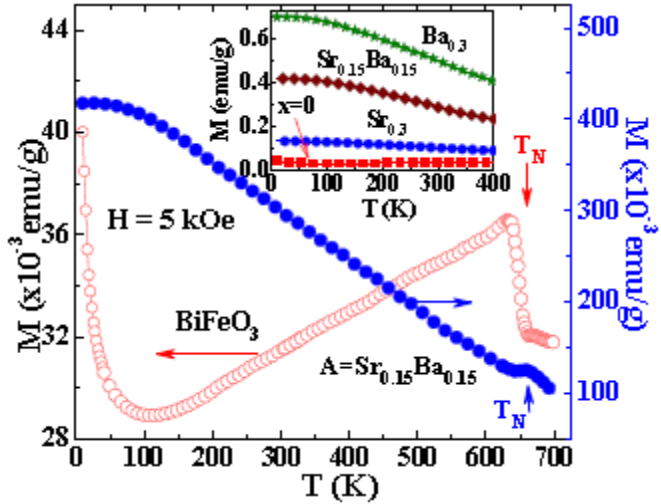

**Fig. 3** The temperature dependences (10 K – 700 K) of magnetization at H = 5 kOe for $BiFeO_3$ and $Bi_{0.7}Sr_{0.15}Ba_{0.15}FeO_3$ compounds. The inset shows the temperature dependences of magnetization (T = 10K – 400 K) at 5 kOe for all the compounds.

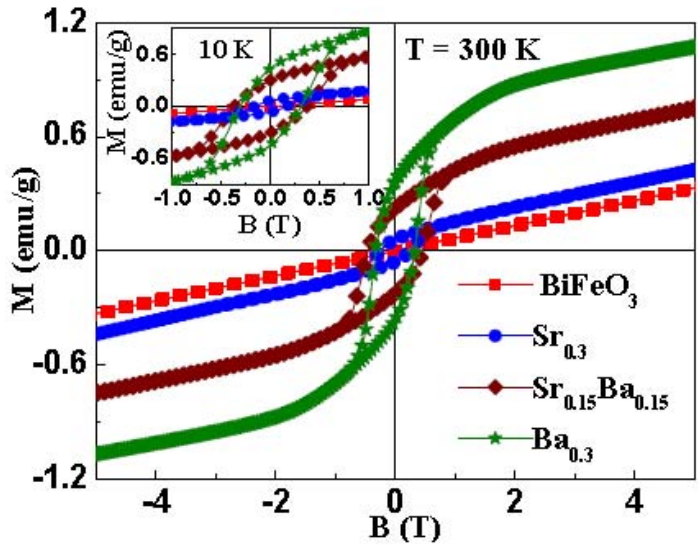

**Fig. 4** Field dependences of magnetization for $Bi_{0.7}(Sr,Ba)_{0.3}FeO_3$ compounds at room temperature. The inset shows the field dependences of magnetization at 10 K.

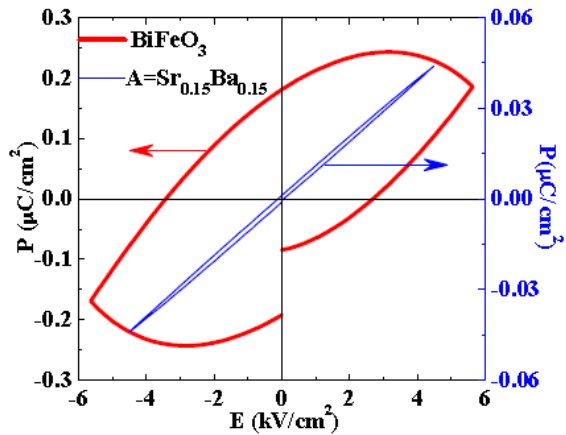

**Fig. 5** *P-E* loops for $BiFeO_3$ and $Bi_{0.7}Sr_{0.15}Ba_{0.15}FeO_3$ compounds at room temperature.

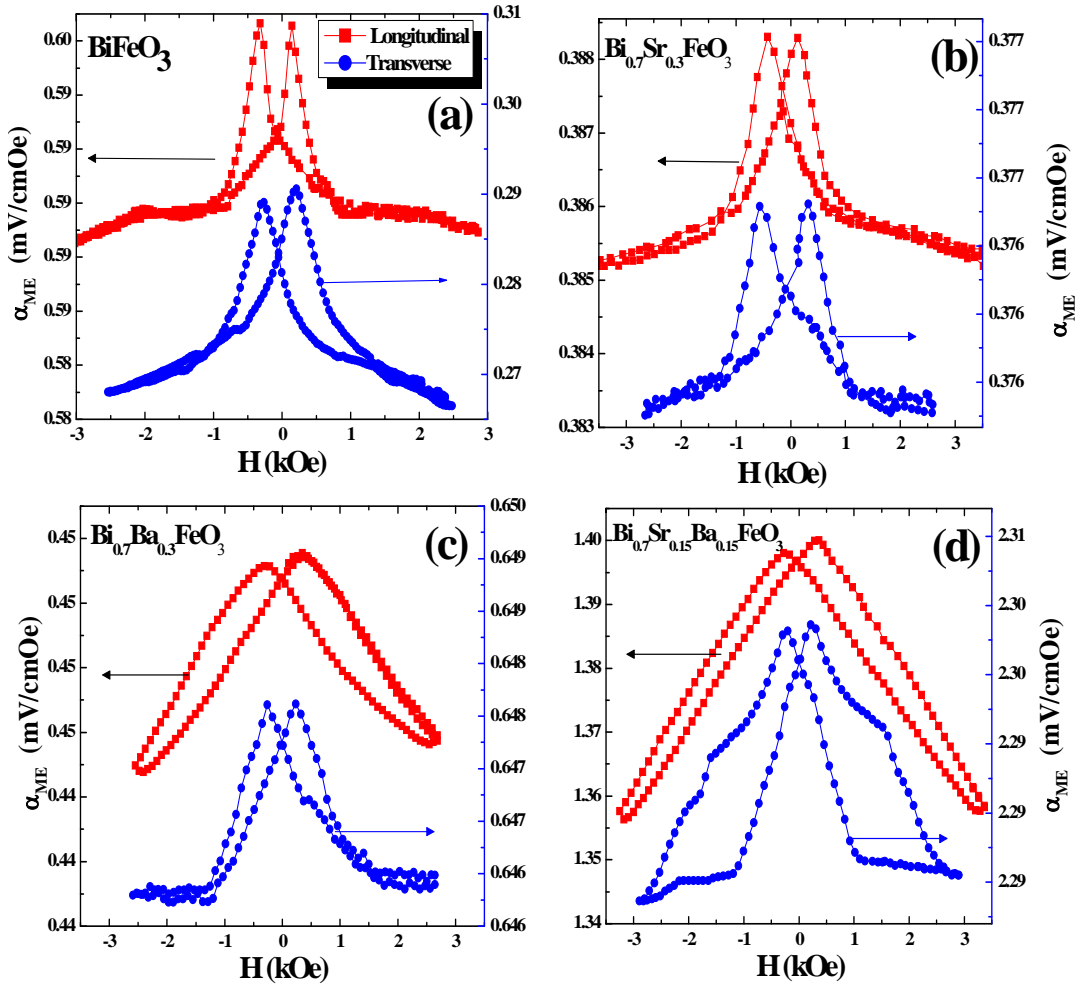

**Fig. 6** Room temperature *dc* bias magnetic field dependence of longitudinal (left scale) and transverse (right scale) magnetoelectric coefficients ($\alpha_{ME}$) for $BiFeO_3$ (x = 0) and $Bi_{0.7}(Sr,Ba)_{0.3}FeO_3$ compounds in ac magnetic field of frequency *f* = 7 kHz.